\documentclass[12pt]{iopart}

\usepackage{iopams}

\usepackage{amsthm}

\usepackage{mathrsfs}
\usepackage{amssymb}

\usepackage{color}

\newtheorem{prop}{Proposition}

\def\be{\begin{equation}}
\def\ee{\end{equation}}
\def\bea{\begin{eqnarray}}
\def\eea{\end{eqnarray}}
\def\bean{\begin{eqnarray*}}
\def\eean{\end{eqnarray*}}

\def\fin{\hfill \rule{2.5mm}{2.5mm}\\ \vspace{0mm}}

\newcommand\ben{\begin{enumerate}}
\newcommand\een{\end{enumerate}}
\newcommand\bit{\begin{itemize}}
\newcommand\eit{\end{itemize}}

\begin{document}

\rapid[Algebraic classification based on the superenergy tensor]{Algebraic classification of the Weyl tensor in higher dimensions based on its ``superenergy'' tensor}

\author{Jos\'e M M Senovilla}
\address{F\'{\i}sica Te\'orica, Universidad del Pa\'{\i}s Vasco, Apartado 644, 48080 Bilbao, Spain}
\eads{\mailto{josemm.senovilla@ehu.es}}

\begin{abstract}
The algebraic classification of the Weyl tensor in arbitrary dimension $n$ is recovered by means of the principal directions of its ``superenergy" tensor. 
This point of view can be helpful in order to compute the Weyl aligned null directions explicitly, and permits to obtain the algebraic type of the Weyl tensor by computing the principal eigenvalue of rank-2 symmetric future tensors. The algebraic types compatible with states of intrinsic gravitational radiation can then be explored. The underlying ideas are general, so that a classification of {\em arbitrary} tensors in general dimension can be achieved.
\end{abstract}

The Petrov classification (e.g. \cite{Exact,PR,Bel2}) of 4-dimensional spacetimes can be reformulated by using the principal directions of the Bel-Robinson tensor. These are the causal vectors whose contraction with the Bel-Robinson tensor vanishes. The underlying ideas go back to \cite{Beltesis,Bel2,D0,L}, are implicit in \cite{PR} and were fully exploited in \cite{Ber} by using spinors. The result follows because the principal directions of the Bel-Robinson tensor coincide with the principal null directions of the Weyl tensor. To summarize, let \cite{Bel,BoS1,S}
\be 
{\cal T}_{\alpha\beta\lambda\mu}=C_{\alpha\rho\lambda\sigma}
C_{\beta}{}^{\rho}{}_{\mu}{}^{\sigma}+C_{\alpha\rho\mu\sigma}
C_{\beta}{}^{\rho}{}_{\lambda}{}^{\sigma}-\frac{1}{8}g_{\alpha\beta}
g_{\lambda\mu}C_{\rho\tau\sigma\nu}C^{\rho\tau\sigma\nu} , \, \, \, \mbox{if $n=4$}
\label{BR*}
\ee
be the Bel-Robinson tensor of a {\em 4-dimensional} spacetime, where $C_{\alpha\rho\lambda\sigma}$ is the Weyl tensor and $g_{\lambda\mu}$ the metric tensor. ${\cal T}_{\alpha\beta\lambda\mu}$ is completely symmetric and traceless. The Petrov classification can be described as follows \cite{Ber}: there exists a null vector $\ell^\mu$ such that
\begin{enumerate}
\item ${\cal T}_{\alpha\beta\lambda\mu}\ell^\mu =0 \Longleftrightarrow $ Petrov type N
\item ${\cal T}_{\alpha\beta\lambda\mu}\ell^\lambda\ell^\mu =0$ but ${\cal T}_{\alpha\beta\lambda\mu}\ell^\mu \neq 0 \Longleftrightarrow $ Petrov type III
\item ${\cal T}_{\alpha\beta\lambda\mu}\ell^\beta\ell^\lambda\ell^\mu =0$ but ${\cal T}_{\alpha\beta\lambda\mu}\ell^\lambda\ell^\mu \neq 0 \Longleftrightarrow $ Petrov type II or D
\item ${\cal T}_{\alpha\beta\lambda\mu}\ell^\alpha\ell^\beta\ell^\lambda\ell^\mu =0$ but ${\cal T}_{\alpha\beta\lambda\mu}\ell^\beta\ell^\lambda\ell^\mu \neq 0 \Longleftrightarrow $ Petrov type I

\vspace{5mm}
Furthermore, one can distinguish types II and D by using
\item there exist two linearly independent null vectors $\ell^\mu$ and $k^\mu$ such that
${\cal T}_{\alpha\beta\lambda\mu}\ell^\beta\ell^\lambda\ell^\mu =0$ and ${\cal T}_{\alpha\beta\lambda\mu}k^\beta k^\lambda k^\mu =0 \Longleftrightarrow $ Petrov type D .
\end{enumerate}
Actually, one can drop the assumption that the vectors are null by just assuming that they are causal (i.e., timelike or null).

In recent years, the algebraic classification of the Weyl tensor in higher dimensions has been constructed based on the existence of WANDs (Weyl aligned null directions) and their alignment order \cite{CMPP,MCPP}, see \cite{C} for a review. Criteria to characterize the (multiple) WANDs, analogous to the classical Bel-Debever criteria \cite{Beltesis,Bel2,D,D1}, have been also recently obtained by Ortaggio \cite{O}. In 4-dimensonal spacetimes, these criteria are known to be equivalent to the number of times one has to contract with the Bel-Robinson tensor, and thereby they provide an alternative view of the Petrov classification. I am going to show that the same happens in arbitrary dimension.

To that end, one only needs the correct generalization of the Bel-Robinson tensor to higher dimensions. As argued in \cite{S}, this is the so-called `superenergy' tensor $T\{C\}$ of the Weyl tensor\footnote{Given any tensor $A$ the super-energy construction \cite{S} is a general method to build a ---essentially unique--- tensor $T\{A\}$ quadratic in $A$ and {\it future}.  Future tensors are those satisfying the dominant property (e.g. (\ref{dp})); see \cite{S} section 4, section 2 in \cite{BS}, or the Appendix in \cite{GS'} for further details. Due to historical reasons \cite{XYZZ,Bel,Bel1,Beltesis,Bel2,D0,D,D1,L,S}, $T\{A\}$ is  called the Ôsuper-energyÕ tensor of $A$.}, whose explicit expression in arbitrary dimension $n$ is
\bea
{\cal T}_{\alpha\beta\lambda\mu}&\equiv &
C_{\alpha\rho\lambda\sigma}
C_{\beta}{}^{\rho}{}_{\mu}{}^{\sigma}
+C_{\alpha\rho\mu\sigma}
C_{\beta}{}^{\rho}{}_{\lambda}{}^{\sigma}
-\frac{1}{2}g_{\alpha\beta}
C_{\rho\tau\lambda\sigma}C^{\rho\tau}{}_{\mu}{}^{\sigma}\nonumber\\
&&-\frac{1}{2}g_{\lambda\mu}
C_{\alpha\rho\sigma\tau}C_{\beta}{}^{\rho\sigma\tau}+
\frac{1}{8}g_{\alpha\beta}g_{\lambda\mu}
C_{\rho\tau\sigma\nu}
C^{\rho\tau\sigma\nu} \label{BR}
\eea
from where one deduces the symmetry properties
\be
{\cal T}_{\alpha\beta\lambda\mu}={\cal T}_{(\alpha\beta)(\lambda\mu)}=
{\cal T}_{\lambda\mu\alpha\beta}.\label{BRsym}
\ee
However, ${\cal T}_{\alpha\beta\lambda\mu}$ is completely symmetric {\em only} in dimensions $n\in \{4,5\}$ \cite{S}.
Similarly, unlike in the case $n=4$, all traces of the tensor (\ref{BR}) are (generically) different from zero for all $n>4$. Expression (\ref{BR}) reduces to the original expression (\ref{BR*}) if $n=4$ due to a 4-dimensional identity \cite{S,EW}.

As any other superenergy tensor, the generalized Bel-Robinson tensor (\ref{BR}) is a future tensor, that is to say, it satisfies the {\em dominant property} \cite{S}
\be
{\cal T}_{\alpha\beta\lambda\mu}v^\alpha_1 v^\beta_2 v^\lambda_3 v^\mu_4 \geq 0, \label{dp}
\ee
for arbitrary future-pointing vectors $v^\alpha_1, v^\beta_2, v^\lambda_3$ and $v^\mu_4$. Inequality (\ref{dp}) is strict if $v^\alpha_1, v^\beta_2, v^\lambda_3$ and $v^\mu_4$ are all timelike \cite{S,BS}. Thus, any possible causal vector $v^\mu$ with the property
\be
{\cal T}_{\alpha\beta\lambda\mu}v^\alpha v^\beta v^\lambda v^\mu =0 \label{pnd}
\ee
must be null. Causal vectors satisfying (\ref{pnd}) define the {\em principal directions} of ${\cal T}_{\alpha\beta\lambda\mu}$ \cite{BS,GS,GS'}. As shown in \cite{PP} with full generality --for the superenergy tensor $T\{A\}$ of any tensor $A$ and in arbitrary dimension---, these are precisely the principal null directions of the Weyl tensor, that is to say, those null vectors satisfying
$$
v_{[\beta}C_{\alpha]\rho\sigma[\lambda}v_{\mu]}v^\rho v^\sigma =0 .
$$
This is, in fact, the characterization of WANDs \cite{MCPP,PrPr}. 

\begin{prop}
At any point of a causally orientable $n$-dimensional Lorentzian manifold where the Weyl tensor does not vanish, a causal vector $\ell^\mu$ satisfies:
\begin{enumerate}
\item ${\cal T}_{\alpha\beta\lambda\mu}\ell^\alpha\ell^\beta\ell^\lambda\ell^\mu =0$ if and only if $\ell_{[\beta}C_{\alpha]\rho\sigma[\lambda}\ell_{\mu]}\ell^\rho \ell^\sigma =0$, that is $\ell^\mu$ is a WAND.
\item  ${\cal T}_{\alpha\beta\lambda\mu}\ell^\alpha\ell^\beta\ell^\lambda =0$ if and only if $\ell_{[\beta}C_{\alpha]\rho[\sigma\lambda}\ell_{\mu]}\ell^\rho=0$.
\item ${\cal T}_{\alpha\beta\lambda\mu}\ell^\alpha\ell^\beta =0$ if and only if $\ell_{[\beta}C_{\alpha]\rho\sigma\lambda}\ell^\rho=0$.
\item ${\cal T}_{\alpha\beta\lambda\mu}\ell^\alpha\ell^\lambda =0$ if and only if $\ell_{[\beta}C_{\alpha\rho][\sigma\lambda}\ell_{\mu]}=0$ \textcolor{blue}{and $C_{\alpha\rho[\sigma\lambda}\ell_{\mu]}\ell^\rho=0$ (which together actually entail $\ell_{[\beta}C_{\alpha]\rho\sigma\lambda}\ell^\rho=0$)}.
\item ${\cal T}_{\alpha\beta\lambda\mu}\ell^\alpha =0$ if and only if $\ell_{[\beta}C_{\alpha\rho]\sigma\lambda}=0$.
\end{enumerate}
In all cases $\ell^\mu$ is necessarily null.
\end{prop}
\proof 
As stated above, the first point (i) is a consequence of a fully general result for superenergy tensors $T\{A\}$ \cite{PP}: once one has sorted out the number of antisymmetric blocks of the seed tensor $A$, the principal directions of $T\{A\}$ are precisely the null directions such that their contraction (or inner product) followed by the exterior product on {\em all} skew-symmetric blocks of $A$ vanishes. However, I am going to include an elementary proof here for the case of the Weyl tensor as this will be illustrative and because it will be useful in the rest of the cases.

Take any null $\ell^\mu$. Contracting with (\ref{BR}) one gets
$$
{\cal T}_{\alpha\beta\lambda\mu}\ell^\alpha\ell^\beta\ell^\lambda\ell^\mu =
2 C_{\rho\sigma}C^{\rho\sigma}
$$
where I have defined
\be
C_{\alpha\beta}(\ell) \equiv C_{\alpha\rho\beta\sigma}\ell^\rho\ell^\sigma .\label{Tll}
\ee
Taking into account the obvious properties $C_{\alpha\beta}=C_{\beta\alpha}$ and $C_{\alpha\beta}\ell^\beta =0$ it follows that $C_{\alpha\beta}C^{\alpha\beta}=0$ is equivalent to (for instance by decomposing on a basis)
\be
C_{\alpha\beta}=\ell_\alpha v_\beta + \ell_\beta v_\alpha \label{1st}
\ee
for some $v_\beta$ (such that $\ell^\rho v_\rho =0$). But this means that $\ell_{[\lambda} C_{\alpha][\beta}\ell_{\mu]} =0$, proving (i). 

To prove (ii), suppose that ${\cal T}_{\alpha\beta\lambda\mu}\ell^\alpha\ell^\beta\ell^\lambda =0$, so that in particular (\ref{1st}) must hold, due to (i). Then a direct computation using (\ref{Tll}) and (\ref{1st}) repeatedly gives
\be
0= {\cal T}_{\alpha\beta\lambda\mu}\ell^\alpha\ell^\beta\ell^\lambda = -\ell_\mu \left(\textcolor{blue}{2} v_\rho v^\rho +\frac{1}{2} C_{\rho\tau\sigma}C^{\rho\tau\sigma} \right) \label{step}
\ee
where I have set
$$
C_{\beta\lambda\mu}(\ell) \equiv C_{\alpha\beta\lambda\mu}\ell^\alpha
$$
with the obvious properties $C_{\beta\lambda\mu}=C_{\beta[\lambda\mu]}$, $C_{[\beta\lambda\mu]}=0$, 
$\ell^\beta C_{\beta\lambda\mu}=0$, $\ell^\lambda C_{\beta\lambda\mu}=C_{\beta\mu}=\ell_\beta v_\mu +\ell_\mu v_\beta$. Choose another null vector $k^\mu$ such that $\ell_\mu k^\mu =1$, define $F_{\lambda\mu}\equiv k^\rho C_{\rho\lambda\mu}=F_{[\lambda\mu]} $ and put a hat on any tensor orthogonal to the timelike plane spanned by $\ell^\mu$ and $k^\mu$. Then a typical decomposition proves 
\be
C_{\beta\lambda\mu}= \ell_\beta F_{\lambda\mu} +\ell_\lambda A_{\beta\mu} - \ell_\mu A_{\beta\lambda} +\hat{C}_{\beta\lambda\mu} \label{Tl}
\ee
with $\ell^\rho A_{\rho\sigma}=k^\rho A_{\rho\sigma}=k^\rho A_{\sigma\rho}=0$, $\ell^\rho A_{\sigma\rho}=-v_{\sigma}+(k^\rho v_\rho)\ell_\sigma$ and $\ell^\rho F_{\rho\sigma}=v_{\sigma}+(k^\rho v_\rho)\ell_\sigma$. A direct calculation shows that (\ref{step}) becomes
$$
\ell_{\mu}\left( -v_\rho v^\rho -\frac{1}{2} \hat{C}_{\rho\tau\sigma}\hat{C}^{\rho\tau\sigma}\right)=0
$$
which implies $\hat{C}_{\rho\tau\sigma}=0$ and $v^\mu =\textcolor{blue}{\frac{1}{2}} A\ell^\mu$. Then, (\ref{1st}) simplifies to $C_{\mu\nu}=A\ell_\mu \ell_\nu $ and (\ref{Tl}) to
\be
C_{\beta\lambda\mu}= \ell_\beta F_{\lambda\mu} +\ell_\lambda \hat{A}_{\beta\mu} - \ell_\mu \hat{A}_{\beta\lambda}  \label{Tl2}
\ee
from where $\ell_{[\alpha}C_{\beta][\lambda\mu}\ell_{\nu]} =0$, proving (ii). Observe that $C_{[\beta\lambda\mu]}=0$ implies now
\be
\hat{F}_{\mu\nu}=2\hat{A}_{[\mu\nu]} \hspace{1cm}
\mbox{where}
 \hspace{1cm} F_{\mu\nu}=\ell_\mu p_\nu -\ell_\nu p_\mu +\hat{F}_{\mu\nu} \label{F=A}
\ee
for some $p^\mu$.

Assume now ${\cal T}_{\alpha\beta\lambda\mu}\ell^\alpha\ell^\beta =0$ so that in particular (\ref{Tl2}) and (\ref{F=A}) hold. Using (\ref{Tl2}) a direct computation gives
$$
0={\cal T}_{\alpha\beta\lambda\mu}\ell^\alpha\ell^\beta =\ell_\lambda \ell_\mu \hat{A}_{\rho\sigma}\hat{A}^{\rho\sigma}
$$
so that $\hat{A}_{\rho\sigma}=0$ hence $C_{\beta\lambda\mu}= \ell_\beta F_{\lambda\mu} $ (with $\hat{F}_{\lambda\mu} =0$). Thus $\ell_{[\alpha}C_{\beta]\lambda\mu}= 0$ proving (iii).

If on the other hand ${\cal T}_{\alpha\beta\lambda\mu}\ell^\alpha\ell^\lambda =0$, and using repeatedly (\ref{Tl2}) and (\ref{F=A}) which remain valid in this situation, the calculation provides
\be
0= {\cal T}_{\alpha\beta\lambda\mu}\ell^\alpha\ell^\lambda = \ell_\beta \ell_\mu \left(\hat{F}_{\rho\sigma}\hat{F}^{\rho\sigma} -\hat{A}_{\rho\sigma}\hat{A}^{\sigma\rho} +\frac{1}{8}C_{\rho\tau\sigma\nu}C^{\rho\tau\sigma\nu}\right)
\label{step2}
\ee
so one needs to compute the last term. At this stage, \textcolor{blue}{by setting $q^\mu \equiv (1/2) (\ell_\rho p^\rho) k^\mu - p^\mu$} the Weyl tensor takes the form
\bean
C_{\alpha\beta\lambda\mu}=(\ell_\alpha k_\beta -\ell_\beta k_\alpha)(\ell_\lambda  \textcolor{blue}{q_\mu} -\ell_\mu  \textcolor{blue}{q_\lambda)}+(\ell_\alpha  \textcolor{blue}{q_\beta} -\ell_\beta  \textcolor{blue}{q_\alpha}) (\ell_\lambda k_\mu -\ell_\mu k_\lambda)\\
\textcolor{blue}{-}(\ell_\alpha k_\beta -\ell_\beta k_\alpha)\hat{F}_{\lambda\mu}\textcolor{blue}{-}\hat{F}_{\alpha\beta}(\ell_\lambda k_\mu -\ell_\mu k_\lambda)+
4k_{[\lambda}\hat{A}_{\mu][\beta}\ell_{\alpha]}+4k_{[\alpha}\hat{A}_{\beta][\mu}\ell_{\lambda]}\\
+2\ell_{[\alpha}\hat{U}_{\beta]\lambda\mu}+2\ell_{[\lambda}\hat{U}_{\mu]\alpha\beta}+
4\ell_{[\alpha}\hat{V}_{\beta][\mu}\ell_{\lambda]}+\hat{C}_{\alpha\beta\lambda\mu}
\eean
for some $\hat{U}_{\beta\lambda\mu}=\hat{U}_{\beta[\lambda\mu]}$, $\hat{V}_{\beta\mu}=\hat{V}_{\mu\beta}$ and $\hat{C}_{\alpha\beta\lambda\mu}$ with the same symmetry properties as the Weyl tensor. It follows that
$$
C_{\alpha\beta\lambda\mu}C^{\alpha\beta\lambda\mu}=4(\ell_\rho p^\rho)^2- \textcolor{blue}{4}\hat{F}_{\rho\sigma}\hat{F}^{\rho\sigma}+8\hat{A}_{\rho\sigma}\hat{A}^{\sigma\rho}  \textcolor{blue}{+} \hat{C}_{\alpha\beta\lambda\mu}\hat{C}^{\alpha\beta\lambda\mu}
$$
so that (\ref{step2}) leads to
$$
 \textcolor{blue}{4(\ell_\rho p^\rho)^2+4\hat{F}_{\rho\sigma}\hat{F}^{\rho\sigma}+\hat{C}_{\alpha\beta\lambda\mu}\hat{C}^{\alpha\beta\lambda\mu}=0}
$$
implying $\ell_\rho p^\rho=0$, \textcolor{blue}{$\hat{F}_{\lambda\mu}=0$} and $\hat{C}_{\alpha\beta\lambda\mu}=0$. \textcolor{blue}{Using the trace-free property of the Weyl tensor, this last result implies also that $\hat{A}_{(\lambda\mu)}=0$, which together with $\hat{F}_{\lambda\mu}=0$ provides $\hat{A}_{\lambda\mu}=0$. All this means $\ell_{[\beta}C_{\alpha\rho][\sigma\lambda}\ell_{\mu]}=0$ (and furthermore $C_{\mu\nu}=0$, as $A=-\ell_\rho p^\rho=0$) plus the condition found previously in (iii), proving (iv). Observe that (iv) always implies (iii).}

Finally, assume that ${\cal T}_{\alpha\beta\lambda\mu}\ell^\alpha =0$ so that everything that has been derived in (iii) and (iv) also holds, that is, $\hat{A}_{\mu\nu}=0$, $\hat{F}_{\mu\nu}=0$, $\ell_\rho p^\rho=0$ (so that $p^\rho=\hat{p}^\rho$) and $\hat{C}_{\alpha\beta\lambda\mu}=0$. One easily gets
$$
0={\cal T}_{\alpha\beta\lambda\mu}\ell^\alpha = -\ell_\beta\ell_\lambda\ell_\mu \left( \hat{p}_\rho \hat{p}^\rho +\textcolor{blue}{\frac{1}{2}}\hat{U}_{\rho\tau\nu}\hat{U}^{\rho\tau\nu}\right)
$$
providing $p^\mu =0$ and $\hat{U}_{\beta\lambda\mu}=0$ so that the Weyl tensor adopts the very simple form $C_{\alpha\beta\lambda\mu}=4\ell_{[\alpha}\hat{V}_{\beta][\mu}\ell_{\lambda]}$ and thus $\ell_{[\nu}C_{\alpha\beta]\lambda\mu}=0$ which ends the proof.
\fin

Combining this proposition with the results in \cite{O} one immediately obtains the algebraic classification of the Weyl tensor using the notation of \cite{CMPP,C}. If $C_{\alpha\beta\lambda\mu}|_x\neq 0$, then its algebraic type can be characterized according to whether there exists a null vector $\ell^\mu$ at $x$ such that:
\begin{itemize}
\item[{\bf N}] $\Longleftrightarrow {\cal T}_{\alpha\beta\lambda\mu}\ell^\alpha =0$. In this case, $\ell^\mu$ is the unique null vector (up to proportionality factors) with this property, and it defines the unique WAND.

\item[\bf{III}] \textcolor{blue}{$\Longleftrightarrow {\cal T}_{\alpha\beta\lambda\mu}\ell^\alpha \ell^\lambda =0$
but ${\cal T}_{\alpha\beta\lambda\mu}\ell^\alpha \neq 0$. Observe that in this case ${\cal T}_{\alpha\beta\lambda\mu}\ell^\alpha \ell^\beta=0$ holds necessarily. The null $\ell^\mu$ with this property is uniquely defined and characterizes the unique {\em multiple} WAND.}


\item[{\bf II} or {\bf D}]  $\Longleftrightarrow {\cal T}_{\alpha\beta\lambda\mu}\ell^\alpha \ell^\beta\ell^\lambda=0$ \textcolor{blue}{but ${\cal T}_{\alpha\beta\lambda\mu}\ell^\alpha \ell^\lambda$ is non-zero for all such $\ell^\mu$.}
\item[{\bf I}]  $\Longleftrightarrow {\cal T}_{\alpha\beta\lambda\mu}\ell^\alpha \ell^\beta\ell^\lambda\ell^\mu= 0$ but ${\cal T}_{\alpha\beta\lambda\mu}\ell^\alpha\ell^\beta \ell^\lambda\neq 0$ for all such $\ell^\mu$.
\item[{\bf G}]  $\Longleftrightarrow {\cal T}_{\alpha\beta\lambda\mu}\ell^\alpha \ell^\beta\ell^\lambda\ell^\mu\neq 0$ for all null $\ell^\mu$.

\vspace{3mm}
In addition, one can distinguish between types {\bf D} and {\bf II} using the following
\item[{\bf D}]  $\Longleftrightarrow$ there exist two linearly independent null vectors $\ell^\mu$ and $k^\mu$ satisfying  ${\cal T}_{\alpha\beta\lambda\mu}\ell^\alpha \ell^\beta\ell^\lambda={\cal T}_{\alpha\beta\lambda\mu}k^\alpha k^\beta k^\lambda = 0$ \textcolor{blue}{(observe that then ${\cal T}_{\alpha\beta\lambda\mu}\ell^\alpha \ell^\lambda$ and ${\cal T}_{\alpha\beta\lambda\mu}k^\alpha k^\lambda$ must be different from zero).}
\end{itemize}
For these cases {\bf D} and {\bf II}, one can further characterize some of their subcases as follows. If ${\cal T}_{\alpha\beta\lambda\mu}\ell^\alpha \ell^\beta=0$, then the types are {\bf II}$_{abd}$ or {\bf D}$_{abd}$, in the latter case ${\cal T}_{\alpha\beta\lambda\mu}k^\alpha k^\beta =0$ actually holds too. 
\textcolor{blue}{Note that the cases with ${\cal T}_{\alpha\beta\lambda\mu}\ell^\alpha \ell^\beta=0$ and ${\cal T}_{\alpha\beta\lambda\mu}k^\alpha k^\lambda=0$ are impossible.}

%
Finally, for types {\bf I}, {\bf II} (\textcolor{blue}{{\bf II}$_{abd}$})
and {\bf III} there exist secondary subtypes {\bf I}$_i$, {\bf II}$_i$ (\textcolor{blue}{{\bf II}$_{iabd}$})
and {\bf III}$_i$ which are simply characterized by the existence of another simple WAND $k^\mu$, that is, a null vector $k^\mu$ which is linearly independent of the given $\ell^\mu$ and such that ${\cal T}_{\alpha\beta\lambda\mu}k^\alpha k^\beta k^\lambda k^\mu= 0$.

One can now explore the algebraic types compatible with a state of intrinsic gravitational radiation at any point $x$. Following Bel \cite{Beltesis,Bel2} this will happen whenever ${\cal T}_{\alpha\beta\lambda\mu}u^\alpha u^\beta u^\lambda$ is a non-zero null vector for all timelike vector $u^\mu$. This definition seems satisfactory in arbitrary dimension because in static cases, when there exists a timelike hypersurface-orthogonal Killing vector $\xi^\mu$, there will never be intrinsic gravitational radiation. To see this, use the result in \cite{LSV} stating that for such $\xi^\mu$, ${\cal T}_{\alpha\beta\lambda\mu}\xi^\alpha\xi^\beta\xi^\lambda = F \xi_\mu$  (in Ricci-flat cases for simplicity). This rules out types {\bf G}, {\bf I}$_i$ and {\bf D} according to \cite{PPO}.

Apart from its intrinsic interest, the previous characterization of the algebraic types of the Weyl tensor can be useful in order to compute the WANDs and to actually classify explicit spacetimes. To take full advantage of this alternative one must use some of the general properties of future tensors,
see \cite{BS,GS'} for the needed details.

Using that ${\cal T}_{\alpha\beta\lambda\mu}$ is a future tensor, first of all one can compute \textcolor{blue}{${\cal T}_{\rho\beta\lambda\mu}{\cal T}^{\rho}{}_{\alpha\tau\nu}$}. If the result is zero, then necessarily ${\cal T}_{\alpha\beta\lambda\mu}=\ell_{\alpha}t_{\beta\lambda\mu}$ for a null $\ell_\mu$ \cite{BS} and thus $\ell^\alpha {\cal T}_{\alpha\beta\lambda\mu}=0$, so that the Petrov type is {\bf N}. Observe that one does not need to know anything about WANDs to check this result. In order to know the unique WAND, simply contract with an arbitrary timelike vector $u^\mu$ thrice: $\ell_\mu ={\cal T}_{\alpha\beta\lambda\mu}u^\alpha u^\beta u^\gamma$. 

\textcolor{blue}{Suppose then that ${\cal T}_{\rho\beta\lambda\mu}{\cal T}^{\rho}{}_{\alpha\tau\nu}$ does not vanish. For {\em any} causal (timelike or null) $w^\mu$ define $\tilde{\cal T}_{\beta\mu}(w) \equiv {\cal T}_{\alpha\beta\lambda\mu}w^\alpha w^\lambda$ which is a symmetric rank-2 future tensor. If $\tilde{\cal T}_{\beta\mu}(w)=0$ then $w^\mu$ is necessarily null and the Petrov type is {\bf III}. If $\tilde{\cal T}_{\beta\mu}(w)\neq 0$, then one looks for its null eigenvectors. In order to know them {\em only the principal eigenvalue}, call it $\lambda$, is needed \cite{BS,GS'}. One can calculate $\lambda$ by classical methods  ---by power iteration and/or solving an algebraic equation. If $\lambda$ is degenerate or corresponds to a double null eigenvector then \cite{BS,GS'} $\tilde{\cal T}_{\beta\mu}\ell^\beta =\lambda \ell_\mu$ for some null $\ell^\mu$, so that $\tilde{\cal T}_{\beta\mu}\ell^\beta\ell^\mu =0$ and one derives ${\cal T}_{\alpha\beta\lambda\mu}\ell^\beta\ell^\mu =0$ leading again to type {\bf III}. The same procedure can be performed with the symmetric rank-2 future tensor ${\cal T}_{\lambda\mu}(u) \equiv {\cal T}_{\alpha\beta\lambda\mu}u^\alpha u^\beta$ for an arbitrary timelike $u^\mu$. If ${\cal T}_{\lambda\mu}\ell^\lambda =\lambda \ell_\mu$ for some null $\ell^\mu$ then ${\cal T}_{\lambda\mu}\ell^\lambda\ell^\mu =0$ and one obtains ${\cal T}_{\alpha\beta\lambda\mu}\ell^\lambda\ell^\mu =0$. All null vectors with this property are those in the eigenspace associated to the principal eigenvalue $\lambda$ of ${\cal T}_{\lambda\mu}(u)$. If there are at least two of them linearly independent, then the Petrov type is {\bf D}$_{abd}$. If there is only one, the Petrov type is {\bf II}$_{abd}$.
Observe that, for the iteration procedure to actually finding these princiapl null eigenspaces, one can use as many ${\cal T}_{\lambda\mu}(u)$, with different $u^\mu$, as desired. Actually, one can even use tensors of type ${\cal T}_{\alpha\beta\lambda\mu}u^\alpha v^\beta$ for any two timelike $u^\mu$ and $v^\mu$.}

If neither ${\cal T}_{\lambda\mu}(u)$ nor $\tilde{\cal T}_{\beta\mu}(u)$ have null eigendirections ---which happens when $\lambda$ corresponds to a non-degenerate eigenvalue with timelike eigenvector --- then one knows that ${\cal T}_{\alpha\beta\lambda\mu}\ell^\lambda\ell^\mu \neq 0$ and ${\cal T}_{\alpha\beta\lambda\mu}\ell^\beta\ell^\mu \neq 0$ for {\em all} possible null $\ell^\mu$. Choose then an {\em arbitrary} null $\ell^\mu$ and construct $T_{\lambda\mu}(\ell )\equiv {\cal T}_{\alpha\beta\lambda\mu}\ell^\alpha\ell^\beta$, which is a non-vanishing future tensor. Computing its principal eigenvalue ---which depends on $\ell^\mu$---, if it happens to be non-degenerate with timelike eigendirection, then there are no WANDs. If it is either degenerate or corresponds to a double null eigenvector but does not (respectively does) vanish, then one must check if the corresponding null eigenvector coincides with $\ell^\mu$ for some choice of the latter, in which case ${\cal T}_{\alpha\beta\lambda\mu}\ell^\alpha\ell^\beta\ell^\lambda\ell^\mu = 0$ (resp. ${\cal T}_{\alpha\beta\lambda\mu}\ell^\alpha\ell^\beta\ell^\mu = 0$). This takes care of types {\bf D}, {\bf II}, {\bf I} and {\bf G}. The calculations in these situations may be long.

It is also possible to provide characterizations of the different types by using tensors and scalars obtained by taking powers of ${\cal T}_{\alpha\beta\lambda\mu}$ and then contracting some or all of the indices. In 4-dimensional spacetimes these results are known but not easy to derive, see \cite{BoS,Ber,FS,FS2}. In higher dimensions they may be even more involved. Nevertheless, they would be very important providing invariant ways of determining the Weyl algebraic types. In a similar vein, the invariants and concomitants written in terms of a general electric-magnetic decomposition could be used, see \cite{XYZZ,Bel,Bel2,BoS,BoS1,S1}.

Notice that the method outlined here applies, mutatis mutandis, to the Riemann tensor, and actually to any double (2,2)-form without the symmetry between pairs of indices. Actually, 
I would like to remark that an algebraic classification of any tensor $A$ can be achieved by the same method, on using its superenergy tensor $T\{A\}$. In general this is a tensor with $2r$ indices, distributed in $r$ symmetric pairs according to the number $r$ of anty-symmetric blocks of indices of the seed tensor $A$ \cite{S,ES}. The number of different algebraic types depends on $r$. As a sufficient illustrative example, consider the case of a rank-2 tensor $A_{\mu\nu}$ (no symmetries assumed), whose superenergy tensor is given by \cite{S}
$$
T_{\alpha\beta\lambda\mu}\{A\}=
A_{\alpha\lambda}A_{\beta\mu}+A_{\beta\lambda}A_{\alpha\mu}-
g_{\alpha\beta}A_{\rho\lambda}A^{\rho}{}_{\mu}-g_{\lambda\mu}A_{\alpha\rho}A_{\beta}{}^{\rho}+
\frac{1}{2}g_{\alpha\beta}g_{\lambda\mu}A_{\rho\sigma}A^{\rho\sigma}
$$
with the property $T_{\alpha\beta\lambda\mu}=T_{(\alpha\beta)(\lambda\mu)}$ (if $A_{\mu\nu}$ happens to be symmetric  then $T_{\alpha\beta\lambda\mu}=T_{\lambda\mu\alpha\beta}$ holds too. In the skew-symmetric case the proper superenergy tensor has only a pair of indices, and the tensor above is a generalization containing it \cite{S2}). An elementary calculation provides the following equivalences (in all cases $\ell^\mu$ is null):
\begin{itemize}
\item $T_{\alpha\beta\lambda\mu}\ell^\alpha\ell^\beta\ell^\lambda\ell^\mu=0 \Longleftrightarrow A_{\mu\nu}\ell^\mu\ell^\nu =0$ 
\item $T_{\alpha\beta\lambda\mu}\ell^\alpha\ell^\beta\ell^\lambda=0 \Longleftrightarrow \ell^\mu A_{\mu[\nu}\ell_{\tau]} =0$ 
\item $T_{\alpha\beta\lambda\mu}\ell^\beta\ell^\lambda\ell^\mu=0 \Longleftrightarrow \ell_{[\mu} A_{\tau]\nu}\ell^{\nu} =0$ 
\item $T_{\alpha\beta\lambda\mu}\ell^\alpha\ell^\beta=0 \Longleftrightarrow \ell^\mu A_{\mu\nu} =0$ 
\item $T_{\alpha\beta\lambda\mu}\ell^\lambda\ell^\mu=0 \Longleftrightarrow A_{\mu\nu}\ell^\nu =0$ 
\item $T_{\alpha\beta\lambda\mu}\ell^\alpha\ell^\lambda=0 \Longleftrightarrow \ell_{[\rho} A_{\mu][\nu}\ell_{\tau]} =0$ 
\item $T_{\alpha\beta\lambda\mu}\ell^\alpha=0 \Longleftrightarrow \ell_{[\tau}A_{\mu]\nu} =0$ 
\item $T_{\alpha\beta\lambda\mu}\ell^\lambda=0 \Longleftrightarrow A_{\mu[\nu} \ell_{\tau]}=0$ 
\end{itemize}
As a final remark, I would like to mention that these classifications can be refined, obtaining more information involving several null directions simultaneously, by considering a generalization called the `mathematical energy tensor'  \cite{S2}, but this is out of the scope of this short communication.

\section*{Acknowledgements}
\textcolor{blue}{I thank Oihane F. Blanco for pointing out some mistakes in the previous (published) version of the manuscript.}
Supported by grants FIS2010-15492 (MICINN) and GIU06/37 (UPV/EHU).

\end{document}